\input epsf                           
\documentstyle[aps]{revtex}
\begin{document}  
\title{Klein paradox and antiparticle}
\author{Guang-jiong Ni\thanks{E-mail: Gjni@fudan.ac.cn}}
\address{Department of Physics, Fudan University, Shanghai, 200433,
  P. R. China}
\author{Weimin Zhou\thanks{E-mail: wz214@is6.nyu.edu} and Jun Yan\thanks{E-mail: jy272@scires.nyu.edu}}
\address{Department of Physics, New York University, 4 Washington
  Place, New York, NY, 10003}

\maketitle

\begin{abstract}
The Klein paradox of Klein-Gordon (KG) equation is discussed to show
that KG equation is self-consistent even at one-particle level and the
wave function for antiparticle is uniquely determined by the
reasonable explanation of Klein paradox. No concept of ``hole'' is needed.
\end{abstract}

\vspace{0.2in}

The Klein paradox of Dirac equation (\cite{Klein} see also
\cite{Bjorken}, \cite{Sakurai}) is of great
historical importance for cognizing the existence of antiparticle of
electron (the positron) and explaining qualitatively the pair creation
process in the collision of particle beam with strongly repulsive
electric field. However, the explanation of this Klein paradox usually
resorted to the concept of ``hole'' in the ``negative-energy electron
sea''. So it was difficult to generalize to the case of Klein-Gordon
(KG) equation where it is hopeless to fill the doubly infinite states
of negative energy. Of course, one would say that this problem had
been solved in the quantum field theory (QFT). But it is still
interesting to see that it can also be understood in the category of
quantum mechanics (QM).

Consider a one-dimensional problem for KG equation:

\begin{equation}
[i\hbar \frac {\partial} {\partial t}-V(x)]^2 \psi=-c^2 \hbar ^2 \frac
{\partial ^2} {\partial x^2} \psi +m^2 c^4 \psi  \label{KG}
\end{equation}

with

\begin{equation}
V(x)=\left\{ \begin{array}{ll}
              0 & \mbox{$x<0$} \\
            V_0 & \mbox{$x>0$}
           \end{array} \right.
\end{equation}

The boundary condition is fixed by incident wave function:

\begin{equation}
\psi_i= a \exp[\frac{i}{\hbar}(px-Et)]  \hspace{0.5in} (x<0) \label{particle}
\end{equation}
with $p>0$, $E=\sqrt{p^2c^2+m^2c^4}>0$.

We expect that the particle wave will be partially reflected at
$x=0$, forming a reflected wave $\psi_r$ together with a transmitted
wave $\psi_t$ as follows:

\begin{eqnarray}
\psi_r & =  b\exp[\frac{i}{\hbar}(-px-Et)]   \hspace{0.5in} (x<0)\\
\psi_t & =  b'\exp[\frac{i}{\hbar}(p'x-Et)] \hspace{0.5in} (x>0)
\end{eqnarray}
where ${p'}^2=(E-V_0)^2/c^2-m^2c^2$.

The continuation condition of wave function ($\psi_i+\psi_r$) with
$\psi_t$ at $x=0$ leads to 

\begin{equation}
\left \{
\begin{array}{ll}
\frac{b}{a} & = \frac{p-p'}{p+p'} \vspace{0.1in}\\
\frac{b'}{a}& = \frac{2p}{p+p'}
\end{array}
\right.
\end{equation}

There are two cases to be discussed:

(1)$E+mc^2>V_0>E$

Since $p'=\sqrt{(V_0-E)^2/c^2-m^2c^2}=iq$ becomes purely imaginary,
the transmitted wave
\begin{equation}
\psi_t=b'\exp[-qx-iEt/\hbar] \hspace{0.5in} (x>0) \label{imag}
\end{equation}
is decreasing exponentially along $x$ axis while the reflectivity of
incident wave equals to $1$:
\begin{equation}
R\equiv\left |\frac{b}{a}\right |^2=\frac{|p-iq|^2}{|p+iq|^2}=1 \label{R1}
\end{equation}

(2) $V_0>E+mc^2$

Since now $p'=\pm \sqrt{(V_0-E)^2/c^2-m^2c^2}$ remains real, the
transmitted wave is oscillating while the reflectivity of incident
wave reads
\begin{equation}
R=\left| \frac{b}{a} \right|^2=\frac{|p-p'|^2}{|p+p'|^2}=\left \{
\begin{array}{ll}
<1 & \mbox{if $p'>0$} \\
>1 & \mbox{if $p'<0$} \label{R2}
\end{array}
\right.
\end{equation}      

While the result of (\ref{imag}) with (\ref{R1}) is as expected, an
satisfying explanation for the prediction (\ref{R2}) is
needed. Especially, we need to know the criterion for the choice of
the sign of $p'$ and what happens when $p'<0$?

For this purpose, we should learn from the important observation by
Feshbach and Villars, who recast the KG equation, Eq. (\ref{KG}), into
two coupled Schr\"odinger equations:

\begin{equation}
\left \{
\begin{array}{ll}
(i\hbar\frac {\partial} {\partial t}-V)\varphi &=mc^2\varphi -\frac{\hbar
^2}{2m}\nabla ^2(\varphi +\chi )  \vspace{0.1in} \\
(i\hbar\frac \partial {\partial t}-V)\chi &=-mc^2\chi +\frac{\hbar
^2}{2m}\nabla ^2(\chi +\varphi ) 
\end{array}
\right. \label{invariant}
\end{equation}
with
\begin{equation}
\left \{
\begin{array}{ll}
\varphi &=\frac{1}{2}[(1-\frac {V}{mc^2})\psi
+i\frac{\hbar}{mc^2}\dot\psi] \vspace{0.1in} \\
\chi &=\frac{1}{2}[(1+\frac V{mc^2})\psi -i\frac{\hbar}{mc^2}\dot\psi]
\end{array}
\right. \label{super}
\end{equation}

Correspondingly, the continuity equation takes the following form, 
\begin{equation}
\frac{\partial \rho }{\partial t}+\nabla \cdot \vec{j}=0 \label{con}
\end{equation}
\begin{equation}
\rho =\frac{i\hbar}{2mc^2}(\psi^*\dot{\psi}-\psi \dot{\psi}^*)-\frac{V}{mc^2}\psi^*\psi=\varphi^*\varphi-\chi^*\chi
\end{equation}
\begin{eqnarray}
\vec{j} &=&\frac{i\hbar}{2m}(\psi \nabla \psi^*-\psi^*\nabla\psi) \nonumber\\
        &=&\frac{i\hbar}{2m}[(\varphi \nabla \varphi^*-\varphi^* \nabla
       \varphi)+(\chi\nabla \chi^*-\chi^*\nabla \chi)+(\varphi \nabla
       \chi^*-\chi^*\nabla \varphi)+(\chi\nabla
       \varphi^*-\varphi^*\nabla \chi)]
\end{eqnarray}  

In the example here, we find for the incident wave ($c=1$):
\begin{equation}
\left \{
\begin{array}{ll}
\varphi_i&=\frac{1}{2}(1+\frac{E}{m})\psi_i \vspace{0.1in} \\ 
\chi_i&=\frac{1}{2}(1-\frac{E}{m})\psi_i
\end{array} \hspace{0.5in} (x<0)
\right.
\end{equation}

\begin{equation}
\rho_i=|\varphi_i|^2-|\chi_i|^2=\frac{E}{m}|a|^2>0
\end{equation}
\begin{equation}
j_i=\frac{p}{m}|a|^2>0
\end{equation}
For the reflected wave, one has
\begin{equation}
\rho_r=\frac{E}{m}|b|^2>0
\end{equation}
\begin{equation}
j_r=-\frac{p}{m}|b|^2<0
\end{equation}
The situation for the transmitted wave is more interesting:
\begin{equation}
\left \{
\begin{array}{ll}
\varphi_t&=\frac{1}{2}(1+\frac{(E-V_0)}{m})\psi_t \vspace{0.1in} \\
\chi_t&=\frac{1}{2}(1-\frac{(E-V_0)}{m})\psi_t
\end{array} \hspace{0.5in} (x>0)
\right.
\end{equation}
\begin{equation}
\rho_t=|\varphi_t|^2-|\chi_t|^2=\frac{(E-V_0)}{m}|b'|^2<0
\end{equation}
\begin{equation}
j_t=\frac{p'}{m}|b'|^2
\end{equation}

It seems quite attractive that we should demand $p'<0$ to get $j_t<0$
in conformity with $\rho_t<0$ and to meet the requirement of
Eq. (\ref{con}) so that 
\begin{equation}
j_i+j_r=j_t
\end{equation}
with $|j_r|>j_i$ ($|b|>|a|$, $ R>1$).

The reason is clear. For an observer located at $x>0$, the energy of
particle in the transmitted wave should be measured with respect to
the local potential $V_0$. In other words, the particle has energy
$E'=E-V_0$ locally. Hence the wave function should be redefined as:
\begin{equation}
\psi_t \rightarrow \tilde{\psi_t}=b'\exp[\frac{i}{\hbar}(p'x-E't)]
\hspace{0.5in} (x>0)
\end{equation}

However, since $E'<0$, from the experimental point of view, the
particle with negative energy behaves as an antiparticle. We should
express its wave function as:
\begin{equation}
\tilde{\psi_t}=b'\exp[-\frac{i}{\hbar}(|p'|x-|E'|t)]
\end{equation}
and claim that the energy and momentum of this antiparticle are
$|E'|>0$ and $|p'|>0$ respectively. It moves to the right though
$p'<0$ and $j_t<0$.

So the above analysis of Klein paradox reveals that the KG equation is
reasonable or self-consistent even at the one-particle level. The
crucial point is looking at its wave function as a coherent
superposition of two parts as shown by Eq. (\ref{super}):
\begin{equation}
\psi=\varphi+\chi
\end{equation}
When $|\varphi|>|\chi|$, it describes a particle like
Eq. (\ref{particle}) with the energy and momentum operators:
\begin{equation}
\hat{E}=i\hbar \frac{\partial}{\partial t}, \hspace{0.5in}
\hat{\vec{p}}=-i\hbar \nabla
\end{equation}

When $|\chi|>|\varphi|$, it describes an antiparticle like:

\begin{equation}
\psi_c\sim\exp[-\frac{i}{\hbar}(\vec{p_c}\cdot \vec{x}-E_ct)] \label{anti}
\end{equation}
with the corresponding operators for antiparticle:
\begin{equation}
\hat{E_c}=-i\hbar \frac{\partial}{\partial t}, \hspace{0.5in}
\hat{\vec{p_c}}=i\hbar \nabla \label{antiop}
\end{equation}
which give $E_c>0$ and $\vec{p_c}$ for $\psi_c$ shown at
Eq. (\ref{anti}). In any case, no concept of ``hole'' is needed.

It is interesting to notice that Eqs. (\ref{anti}) and (\ref{antiop})
were pointed out long ago by Schwinger \cite{Schwinger}, Konopinski,
and Mahmaud \cite{Konopinski}, and
even earlier (in the Green function or propagator of QFT) by
St\"uckelberg \cite{Stuckelberg} and Feynman \cite{Feynman} essentially.

However, if we accept the above point of view,it will have
far-reaching consequence. A particle is always not pure, it always
comprises two components, $\varphi$ and $\chi$. In the equation
governing its motion, $\varphi$ and $\chi$ are always coupled together
with the symmetry under the transformation ($\vec{x}\to
-\vec{x}, t \to -t$) and
\begin{eqnarray}
\varphi(-\vec{x},-t)\to\chi(\vec{x},t) \nonumber \\
V(-\vec{x},-t)\to-V(\vec{x},t) \label{trans}
\end{eqnarray}
as shown at Eq. (\ref{invariant}) as a special case.

But we wish to stress that Eq. (\ref{trans}) is a basic symmetry,
which should be raised as a general postulate in relativistic quantum
mechanics as well as in QFT \cite{Ni}. It can also be served as a
starting point
to understand the essence of special relativity. (\cite{NiChen}, \cite{NZY})

Finally, it is interesting to add that another paradox in physics, the
original version of EPR paradox \cite{Einstein} also raised a very acute
question in
quantum mechanics. Its reasonable explanation \cite{NiGuan} leads precisely to the
same conclusion as that in this paper, i.e., the necessity of
existence of antiparticle with its wave function shown as
Eq.(\ref{anti}).

This work was supported in part by the NSF of China. 

\end{document}